# Synthesis and Characterization of NiCoMn MOFs for Wastewater Treatment

*Muhammad Farhan[1], and Osama Aziz[1]*

[1]Faculty of Materials and Chemical Engineering, Ghulam Ishaq Khan Institute of Engineering Sciences and Technology (GIKI), Topi- 23640, Pakistan
E-mail: u2021371@giki.edu.pk

**Abstract:** Water pollution has become a global problem. Sources of wastewater majorly include industrial and commercial sectors. To cater the exponential increase in clean water, efficient technologies are needed to treat wastewater. Several techniques such as redox reactions, membrane filtrations, mechanical processes, chemical treatment and adsorption techniques have been employed. However their cost and effectiveness is still a major problem. In this study we employed an effective wastewater treatment technique by synthesizing NiCoMn MOFs using a simple hydrothermal technique, and characterized the properties using XRD and SEM for their possible characteristics. XRD analysis confirmed the successful synthesis of NiCoMn MOFs. Sufficient information regarding the surface morphology and topology was given by the SEM analysis which proved a nanoporous structure with high surface area effective for adsorption and oxidative catalysis of contaminants in wastewater. Moreover, a high electrostatic attraction between the MOFs was observed which could attract oppositely charged contaminants. The results showed a high potential for the synthesized NiCoMn MOFs for wastewater treatment applications.

## 1. Introduction

Globally, water pollution in urban aquatic habitats has become a major problem. When contaminants are released into water bodies without being properly treated, there is serious pollution and health danger. The most common causes of water contamination include pesticide use, oil spills, nuclear waste leaks, resource extraction, and home and industrial waste. Environmental pollution, especially in aquatic bodies, has reached hazardous levels, according to scientific research [1 - 5]. Since water is necessary for life, protecting it is crucial. There is more demand for water resources than there is population expansion, which puts more strain on water treatment systems. As a result, the demand for affordable water treatment technology is rising. Wastewater treatment involves a variety of techniques, such as advanced oxidation processes, adsorption, oxidation-reduction, membrane filtering, chemical treatment, mechanical processes, and incineration. Adsorption is one of these techniques that stands out as a potential and successful method for treating water. When it comes to wastewater treatment, porous nanoparticles are efficient adsorbents. Zeolites and activated carbon are used to clean wastewater; nevertheless, they have limitations, including small surface surfaces and restricted functionality [6]. Improved adsorbents are needed for the treatment of industrial wastewater. Metal organic frameworks (MOFs) represent a new class of porous nanomaterials with a wide range of applications in the treatment of wastewater.

MOFs have attained a lot of attention throughout the last 50 years. Their porosity allows guest molecules to enter the structure, is one of the important characteristics. Their porous nature is because of the formation of coordinate bonds between organic ligands and metal ions. In line with Yaghi et al. [7]. Due to their porous nature, it has a wide range of applications in fields like, medication delivery, bioreactors, microelectronics, optics, and gas adsorption. Its pore size is as tiny as 2 nm, MOFs are suitable for small molecules [8]. Diffusion, ultrasound-assisted and microwave-assisted approaches are used for the synthesis of MOFs. One of the challenge is that the above mentioned methods are highly expensive and intricate [9].

In this work, we synthesized NiCoMn MOFs using a straightforward hydrothermal synthesis approach, paying close attention to the synthesis conditions and composition. The synthesized MOFs were then characterized through XRD and SEM in order to assess for the wastewater treatment applications.

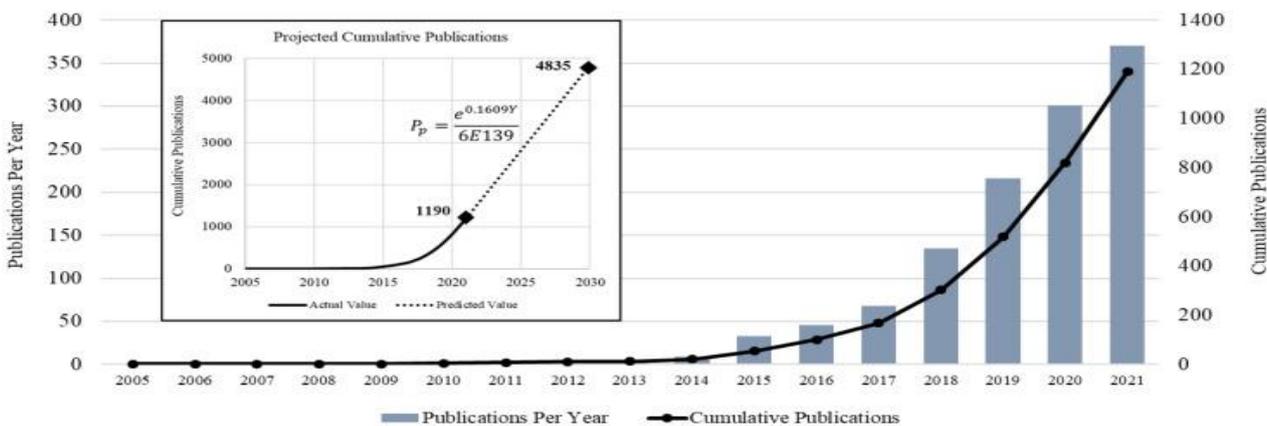

**Fig. 1.** Development of Research in MOF for Applications in Wastewater [10]. Treatment.

## 2. Materials and experimental

### *2.1 Materials*

The materials used for this work were cobalt (II) nitrate hexahydrate (Co $(NO_3)_2 \cdot 6H_2O$), manganese (II) nitrate bi hydrate (Mg $(NO_3)_2 \cdot 2H_2O$), nickel (II) nitrate hexahydrate (Ni $(NO_3)_2 \cdot 6H_2O$) as the metal precursors. For the purpose of ligand, terephthalic acid ($C_8H_6O_4$), was used. N, N dimethylformamide (DMF) was used as a solvent. All of these chemicals were available in lab and were obtained from Sigma Aldrich.

## 2.2 Synthesis of NiCoMn MOFs

A computerized mass balance was used to precisely weigh 0.29103 grams of cobalt (II) hexahydrate, 0.25101 grams of manganese (II) nickel tetra hydrate, and 0.29081 grams of nickel (II) nitrate hexahydrate, which is equivalent to 7 mmol, all of which were added to a 100-milliliter beaker. Then, as a ligand, 0.16613 grams of terephthalic acid, or 5 mmol, was added. Dimethylformamide (DMF), 60 milliliters, was added to the beaker as the solvent to help the reaction along. To guarantee that the components were evenly distributed, the resultant mixture was stirred for 30 minutes at 600 revolutions per minute (rpm) using a magnetic stirrer. After the stirring was finished, the mixture was poured into a 100-milliliter autoclave lined with Teflon and heated to 165°C for 24 hours. Following the completion of the stirring, the mixture was transferred into a 100 cc Teflon-lined autoclave and heated using the hydrothermal synthesis method to 165°C for an entire day. After hydrothermal synthesis, the solution was divided into four test tubes, each holding 14 ml of material and centrifuged for 10 minutes for 3 cycles. The test tubes were rinsed with DI water and ethanol and finally dried in an oven to get the powders.

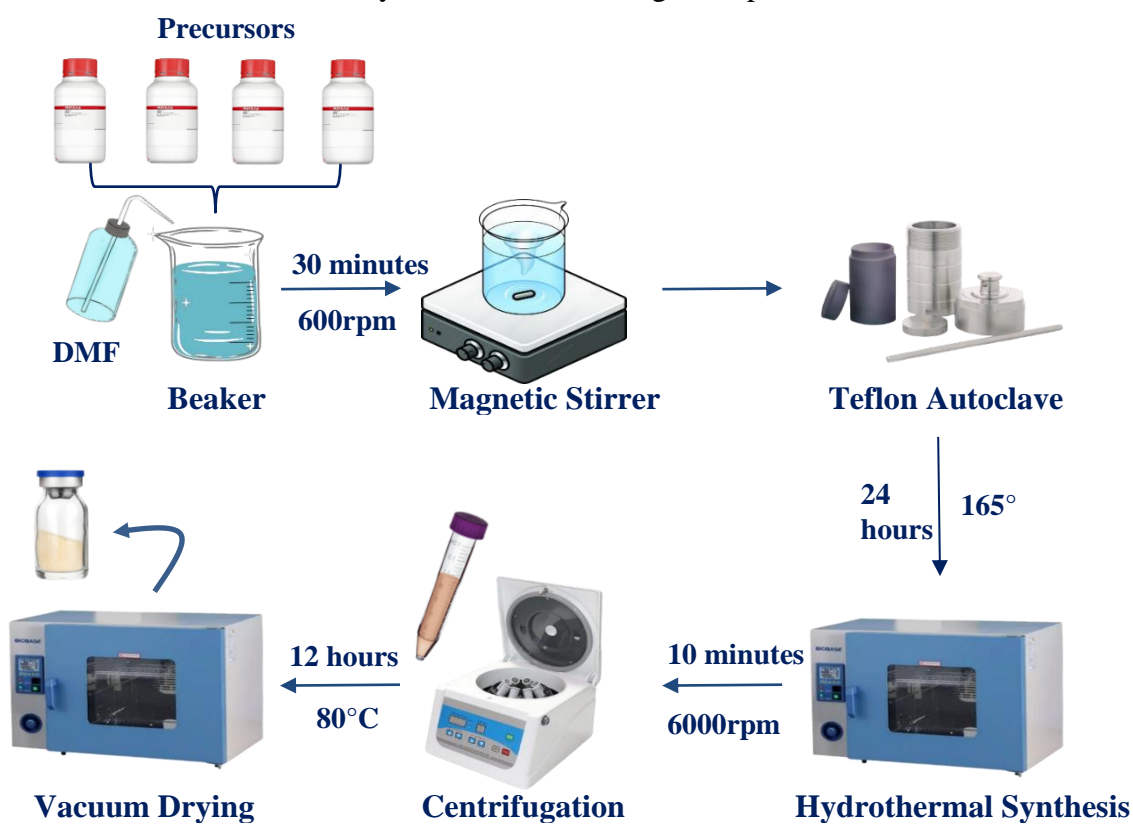

**Fig. 2.** Step-by-step procedure for the synthesis of NiCoMn MOFs.

## 2.3 Characterization techniques

X-ray diffraction machine from Proto was used to analyze the crystal structure and phase composition of the elements present in the sample. Its diffraction angle (2 theta) was set from 10 degrees to 90 degrees having a Cu-Kα radiation with wavelength 1.54 nm. A spectrum was observed for the intensity in au against diffraction angle. A scanning-electron microscope (SEM) with built in Energy dispersive X-ray spectroscopy (EDX) from ZEISS helped in studying the morphology and composition of the material. Since we were limited to only these two techniques, the adsorption related properties for WWT were assessed by admirable work done by earlier scientists.

## 3. Results and discussion

### 3.1 XRD analysis

The crystalline configurations of NiCoMn MOFs was analyzed through XRD examination, as shown in the Fig. 2, to verify the successful fabrication of NiCoMn MOF. Distinctive diffraction peaks of Ni at $2\theta = 44.50°$ and $2\theta = 39.50°$ are observed with reference to JCPDF File 45-1027. Similarly, Co (FCC) at shows peaks at $2\theta = 42.50°$ (111) and $2\theta = 51.70°$ (200). Lastly, Mn peaks appear at $2\theta = 23.90°$ and $2\theta = 31.00°$.

This spectrum aligns well with previous literature [27, 28]. Consequently, the XRD pattern of NiCoMn MOFs confirm the presence of characteristic peaks for Ni, Co, and Mn, thus validating the successful surface synthesis of NiCoMn MOFs.

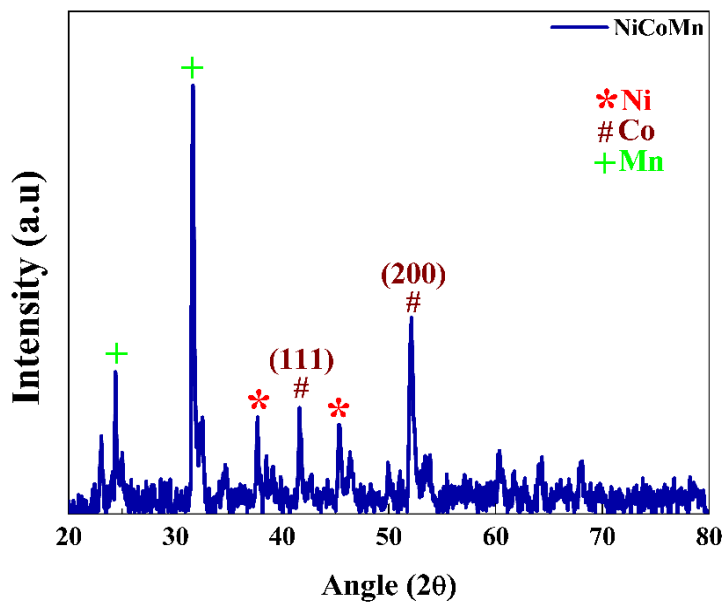

**Fig. 3.** XRD patterns of NiCoMn MOFs.

### 3.2 SEM and EDX analysis

SEM pictures of NiCoMn MOFs depict the topology and surface morphology of the framework. The SEM analysis aimed to assess the adhesion and organization of the trivalent metal organic framework. As illustrated in Fig. 4, images a, and b reveal a porous structure which aligns well with the application requirement [11, 12]. The high surface area as shown in the images is an indication that these transition MOFs can have a good oxidative catalytic behavior in aqueous solutions and can breakdown highly acidic and undesired particles in wastewater effectively [13].

The pictures show that the MOFs are aggregated hence it's difficult to observe them separately, however in the vicinity of the aggregates the structure of MOF can be seen. The agglomerates were formed since we did not use any ultra-sonication technique during synthesis therefore we can conclude that for better synthesis of MOFs ultra-sonication is crucial. This also gives us an idea on high electrostatic interactions of the MOFs, this high adsorption can be useful in removing contaminants from water [14].

Moreover, Fig. 3, (a1-a4) shows the EDX mapping which illustrates the presence of the constituent elements Ni, Co, Mn, and O respectively. This gives us insights into the chemical structure of the NiCoMn Metal Organic Framework (MOF) specifically that all the elements were distributed evenly throughout the framework.

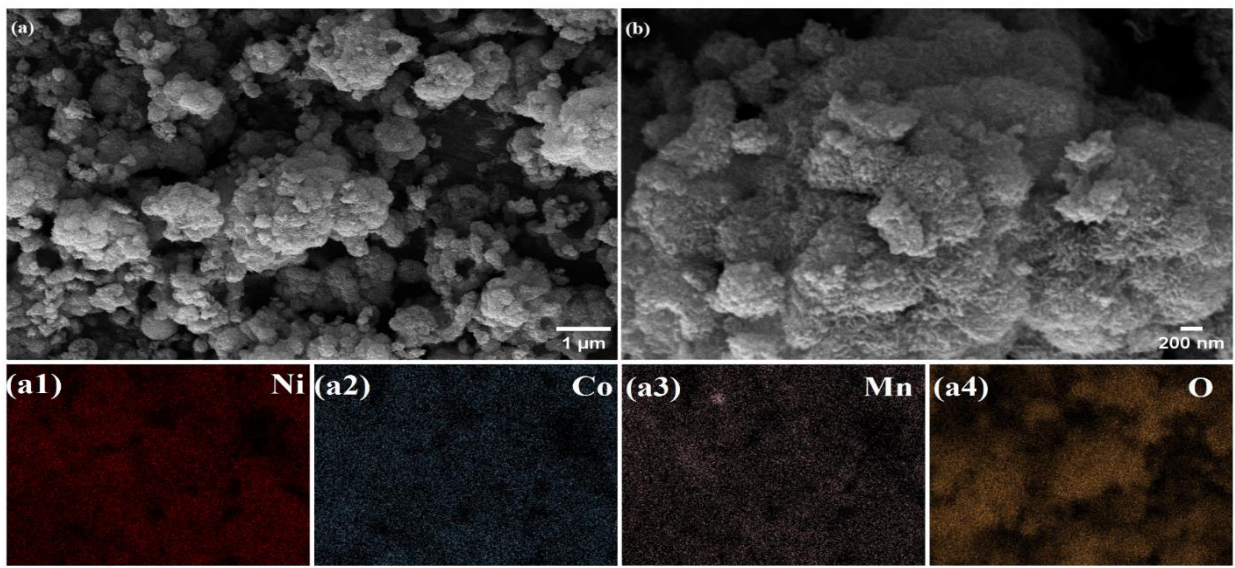

**Fig. 4.** SEM images of (a, b) NiCoMn MOFs at 1 µm and 200 nm. EDX mapping (a1 – a4) illustrates the presence of Ni, Co, Mn and O elements.

## 4. Possible characteristics

### *4.1 Adsorption process*

Adsorption is the transfer of fluid components to a solid material's surface, used for compound separation and wastewater treatment. It is cost-effective and relies on the adsorbent's area of contact. Fig. 5 represents a schematic to show a typical example of how this adsorption will occur. The surface area of the adsorbent has a major influence on the amount of adsorption; tiny particles with high porosity are needed for industrial applications. The adsorbent's characteristics and chemical makeup also have a significant impact on how much adsorption it can hold. In numerous sectors, materials possessing certain functional groups or a strong affinity for target molecules are frequently chosen for effective adsorption procedures. Therefore, choosing the best adsorbent for a given application requires taking into account a variety of factors, including surface area, porosity, and chemical characteristics.

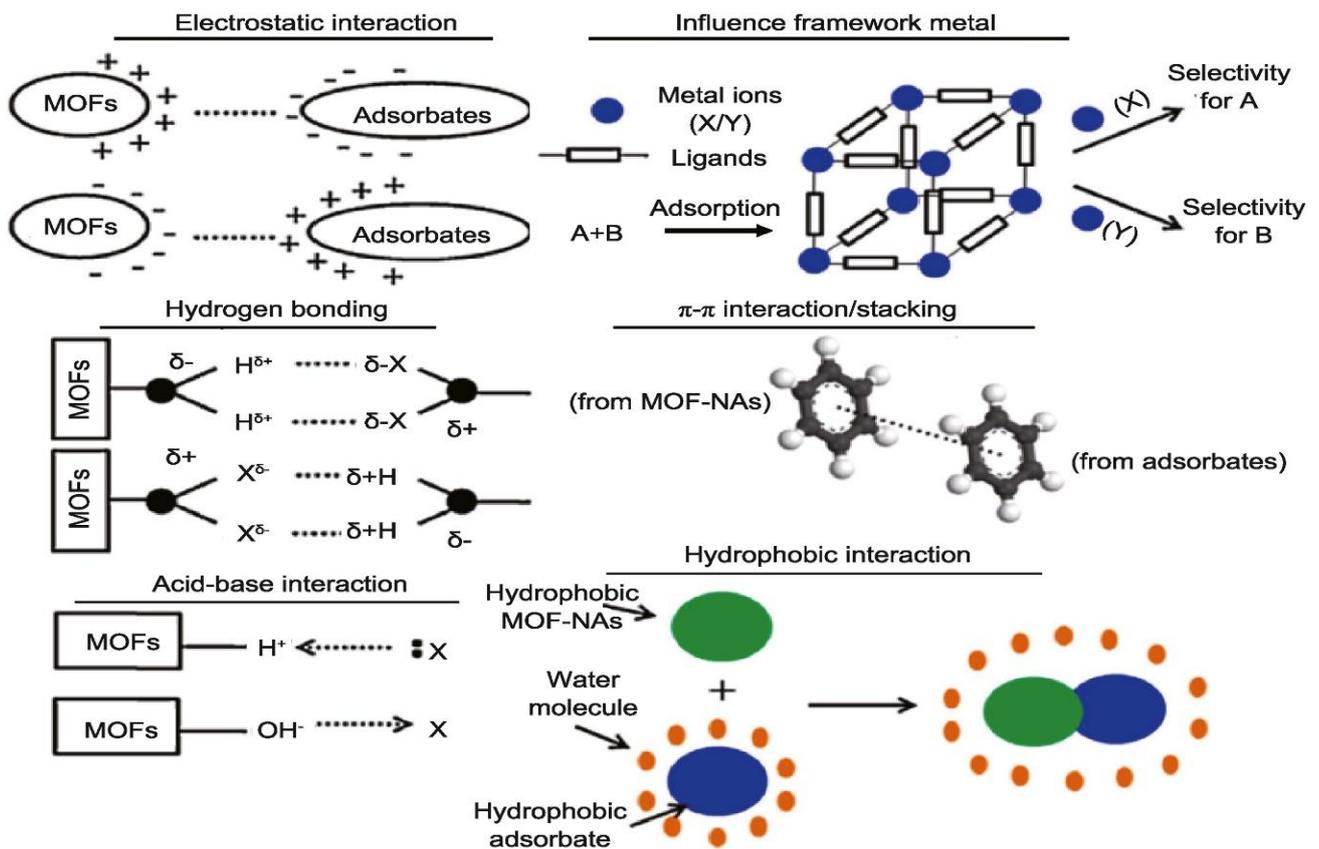

**Fig. 5.** A schematic representation of possible pollutants' adsorption pathways on MOFs [15].

The primary processes in the removal of hazardous compounds using MOFs are double bond interactions, hydrogen bonding, chemical reactions, and opposite charge attractions. Interactions between oppositely charged MOFs (adsorbents) and the surface charges of impurities (adsorbates) are crucial to the adsorption process. Surface charges that result in gain or loss of electron intensify the attraction between MOFs and waste agents [16].

## *4.2 Nanofiltration*

Wastewater contaminants are commonly treated using filter membranes. The top surface's pore size can be used to categorize them. Due to its many advantages, including low energy consumption, great efficiency, and affordability, NF is frequently employed in ecological domains. OSN is a technique that increases the use of membranes and separates organic liquids. There is recognition for the TFC polyamide membrane's energy efficiency [17]. Properties are improved by TFN membranes, and MOFs have the potential for TFN development [18]. NF technology has improved in the purification of wastewater. By merging MOFs and NF membranes, composite membranes with improved WWT properties are created. An $Al_2O_3$ ZIF-300 membrane is used to cleanse water in order to extract heavy metal ions [19].

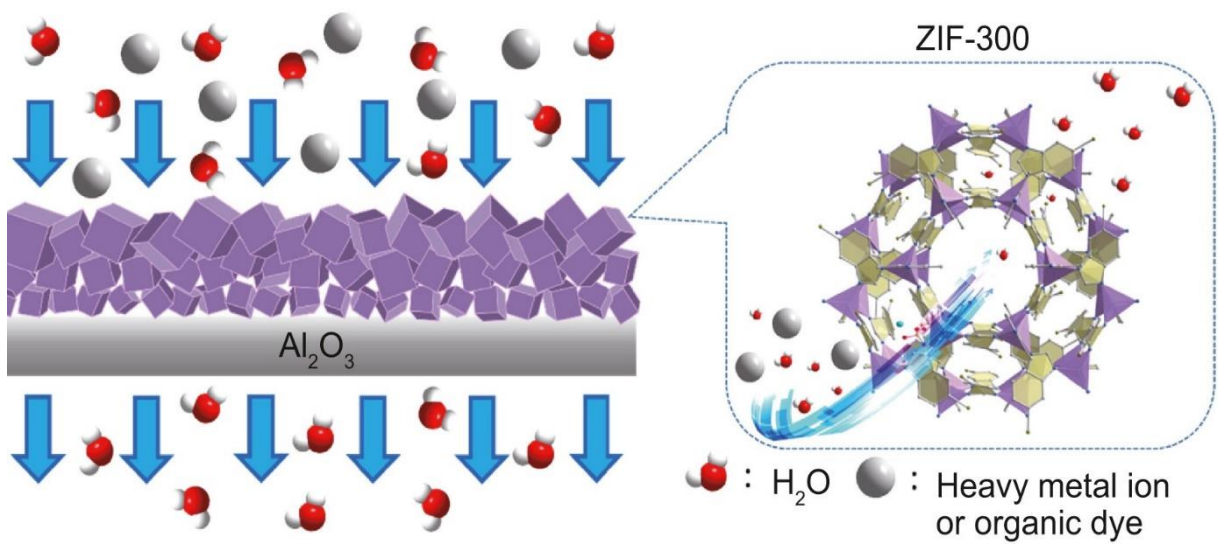

**Fig. 6.** A ZIF-300 membrane eliminating heavy metal ions from wastewater [19].

### *4.3 Reverse osmosis*

Seawater desalination is the main use for RO. Furthermore, it is often employed to eliminate organic pollutants from wastewater [20]. Modified MOF increases filtration effectiveness and lowers energy consumption in RO membranes. $Cu_3(BCT)_2$ MOF treated with acid produced a novel RO membrane that performed better [21]. It demonstrated 96 % better salt interception and 33 % more water flow. Because of its improved hydrophilicity and porous structure, the modified membrane exhibited enhanced anti-fouling capabilities [22, 23]. Certain MOF nanoparticles were combined with TFN membrane to create a high-performing RO composite membrane. For instance, adding ZIF-8 to TFN's PA layer greatly improved desalination performance. While keeping admirable NaCl interception, the 0.4 % ZIF-8 embedding in the TFN membrane enhanced water permeability by 162 % over the original polyamide membranes [160]. Particles ZIF-8, 50 nm in size were found to have the best dispersion in the PA layer, indicating that TFN RO membranes could benefit from their use. MIL-101(Cr) NPs improved water flux in TFN membranes by 44 % without sacrificing sodium chloride rejection. The material MIL-101(Cr), has a large SA and water absorption capacity, provides a channel for the passage of water molecules. Excellent RO composite membranes were also produced by adding UiO-66 and MIL-125 as TFN membrane fillers [24]. UiO-66-NH2 membrane is a promising RO membrane for desalination because it can improve cation selectivity when used on an alumina surface.

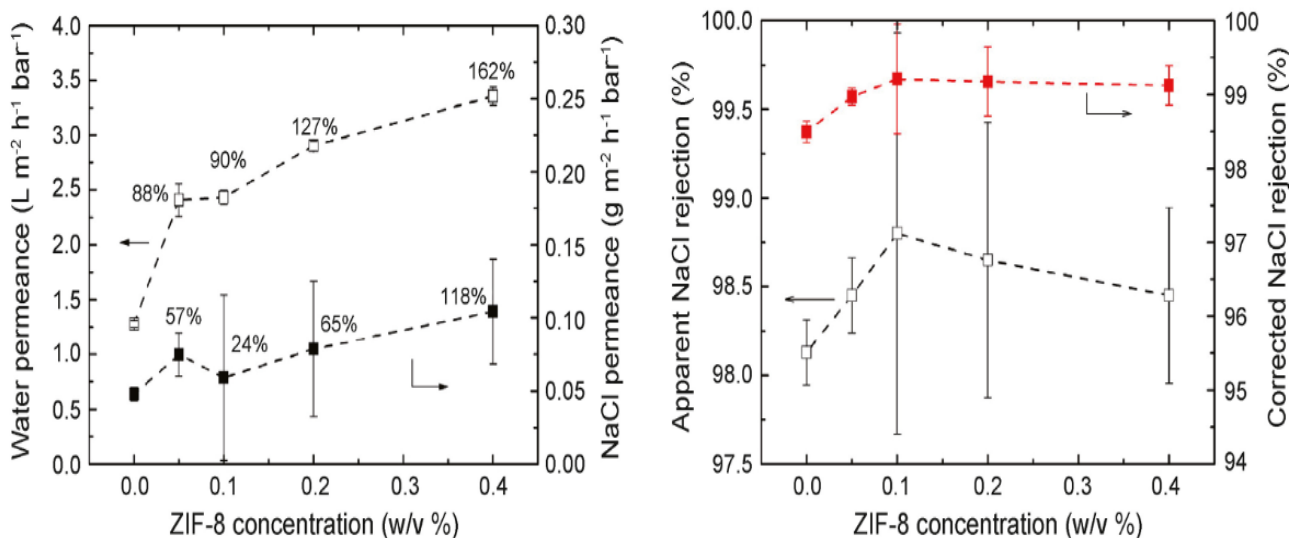
**Fig. 7.** NaCl rejection of the TFN membrane, both apparent and corrected and water presence [21].

## 5. Conclusion

Vast dangerous chemicals released from home, industrial, agricultural, and pharmaceutical sources pollute natural water sources, seriously damaging the ecology. MOF-derived materials' vast contact area and tunable pore size make them useful catalysts and adsorbents in a variety of wastewater pollution removal procedures. Because of their extraordinary separation efficiency, flexibility in size and composition, and other benefits, MOF-based materials can be used to effectively remove different contaminants from wastewater. Therefore, in both academic and industrial settings, MOF-based compounds provide good precursors for wastewater treatment [25].

## 6. Acknowledgements

The authors acknowledge the Higher Education Commission of Pakistan (HEC) and Ghulam Ishaq Khan Institute of Engineering Sciences and Technology for their support on conducting the synthesis and characterization through the lab equipment.